\documentclass[11pt,titlepage]{article}
\usepackage[margin=1.2in]{geometry}
\usepackage{amsmath, amsbsy, amsthm, array, mathrsfs}
\usepackage{graphics}
\usepackage{physics}
\usepackage{epsfig, amssymb,latexsym,verbatim}
\usepackage{graphicx,epstopdf}
\usepackage{color}
\usepackage{dsfont, bbm}
\usepackage{relsize}
\usepackage{subcaption}

\newcommand{\R}{\mathbb{R}}

\newcommand{\noin}{\noindent}
\newcommand{\bee}{\begin{eqnarray*}}
\newcommand{\ene}{\end{eqnarray*}}
\newcommand{\bec}{\begin{center}}
\newcommand{\enc}{\end{center}}
\newcommand{\be}{\begin{equation}}
\newcommand{\ee}{\end{equation}}

\newcommand{\mb}{\mathbf}
\newcommand{\bs}{\boldsymbol}
\newcommand{\tb}{\textbf}
\newcommand{\pend}{$\square$}
\newcommand{\vs}{\vskip 3mm}
\newcommand{\bi}{\begin{itemize}}
\newcommand{\ei}{\end{itemize}}

\begin{document}
\baselineskip 3.2ex

\title{\LARGE Weighted least squares regression with the best robustness and  high computability} 
\vs
\vs
\author{ {\sc Yijun Zuo and Hanwen Zuo}\\[2ex]
         {\small {\em   Department of Statistics and Probability and  Department of Computer Science} }\\[.5ex]
         {\small Michigan State University, East Lansing, MI 48824, USA} \\[2ex]
         {\small zuo@msu.edu and  zuohanwe@msu.edu}\\[6ex]
     }
 \date{\today}
\maketitle

\vskip 3mm
{\small

\begin{abstract}

A novel regression method is introduced and studied. The procedure weights squared-residuals based on their magnitude. Unlike the classic least squares which treats every squared-residual equally important,
the new procedure exponentially down-weights squared-residuals that lie far away from the cloud of all residuals and assigns a constant weight (one) to squared-residuals that lie close to the center of squared-residual cloud.
\vs
The new procedure can keep a good balance between robustness and efficiency, it possesses the highest breakdown point robustness for any regression equivariant procedure, much more robust than the classic least squares, yet much more efficient than the benchmark of robust method, 
the least trimmed squares (LTS) of Rousseeuw (1984). 
 With a smooth weight function, the new procedure could be computed
very fast by fist-order (first-derivative) method and second-order (second-derivative) method.
Assertions and other theoretical findings are verified in simulated and real data examples.

\vs

\bigskip
\par

\noindent{\bf Key words and phrase:} weighted least squares, robustness, efficiency,  computability, finite sample breakdown point.
\bigskip
\par
\noindent {\bf Running title:} Weighted least squares.
\end{abstract}
}
\setcounter{page}{3}

\section {Introduction}
In the classical regression analysis, we assume that there is a relationship for a given data set $\{(\bs{x}^{\top}_i, y_i)^{\top}, i\in \{1,2, \cdots, n\}\}$: 
\be
y_i=(1,\bs{x}^{\top}_i)\bs{\beta}_0+{e}_i,~~ i\in\{1,\cdots, n\}  \label{model.eqn}
\ee
where $y_i\in \R^1$, ${\top}$ stands for the transpose, $\bs{\beta}_0=(\beta_{01}, \cdots, \beta_{0p})^{\top}$ (the true unknown parameter) in $\R^p$ and~ $\bs{x_i}=(x_{i1},\cdots, x_{i(p-1)})^{\top}$ in $\R^{p-1}$ ($p\geq 2$), $e_i\in \R^1$ is called an error term (or random fluctuation/disturbances, which is usually assumed to have zero mean and variance $\sigma^2$ in classic regression theory). That is, $\beta_{01}$ is the intercept term of the model. Write $\bs{w}_i=(1,\bs{x}'_i)^{\top}$, then one has $y_i=\bs{w}^{\top}_i\bs{\beta}_0+e_i$, which will be used interchangeably with (\ref{model.eqn}).
\vs
One wants to estimate the $\bs{\beta}_0$ based on a given sample $\mb{z}^{(n)}
:=\{(\bs{x}^{\top}_i, y_i)^{\top}, i\in\{1,\cdots, n\}\}$ from the model $y=(1,\bs{x}^{\top})\bs{\beta}_0+e$.
 Call the difference between $y_i$ and $\bs{w^{\top}_i}{\bs{\beta}}$
 the ith residual, $r_i(\bs{\beta})$, for a candidate coefficient vector $\bs{\beta}$ (which is often suppressed). That is,
\be r_i:={r}_i(\bs{\beta})=y_i-\bs{w^{\top}_i}{\bs{\beta}}.\label{residual.eqn}
\ee
To estimate $\bs{\beta}_0$, the classic \emph{least squares} (LS) minimizes the sum of squares of residuals,
$$\widehat{\bs{\beta}}_{ls}=\arg\min_{\bs{\beta}\in\R^p} \sum_{i=1}^n r^2_i. $$
Alternatively, one can replace the square above by the absolute value to obtain the least absolute deviations  estimator (aka, $L_1$ estimator, in contrast to the $L_2$ (LS) estimator).
\vs
The LS estimator is  very popular  in practice across a broader spectrum of disciplines due to its great computability and optimal properties when the error $e_i$s are i.i.d. follows a normal ${N}(\bs{0},\sigma^2)$ distribution.
It, however, can behave badly when the error distribution is slightly departed from the normal distribution,
particularly when the errors are heavy-tailed or contain outliers.
\vs
Robust alternatives to the $\widehat{\bs{\beta}}_{ls}$ abound in the literature for a long time. The most popular ones are  M-estimators \cite{H64}, 
the least median squares (LMS) and least trimmed squares (LTS) estimators \cite{R84}, 
  S-estimators  \cite{RY84}, 
   MM-estimators \cite{Y87}, 
   $\tau$-estimators \cite{YZ88}, 
     maximum depth estimators (\cite{RH99}, 
     \cite{Z21a, Z21b, Z21c}), and the lately the least squares of trimmed residuals (LST) regression \cite{ZZ23},  among others. 
     For more related discussions, see, Sections 1.2 and 4.4 of \cite{RL87}, 
     and  Section 5.14 of \cite{MMY06}.
\vs
Robust methods that have a high breakdown point are usually computationally intensive and with a non-differentiable objective function (e.g., LMS, LTS, and LST).
In this article, we will introduce a smooth and differentiable objective function that greatly facilitates the computation of the underlying estimator.
We introduce a new class of alternatives for robust regression,  
weighted least squares (WLS) estimators {$\boldsymbol{\hat \beta}_{wls}$}:
\be
\widehat{\bs{\beta}}_{wls}=\arg\min_{\bs{\beta}\in \R^p} \sum_i^n w_ir^2_i(\bs{\beta}),\label{lws.equ}
\ee
where
$w_i$ is the weight associated with $r_i$ with
a fundamental feature: it assigns equal weight to all $r^2_i$ that are small (no greater than a cut-off value)
and exponentially down-weights (penalizes) the large  ones (when $r^2_i$s are greater than the cut-off value). 
\vs
Weighted least squares estimation has been proposed and discussed in the literature, see, e.g., section 2.1 of \cite{RL87}, section 5.11 of \cite{MMY06}, but not the 
weighted procedure first proposed in this article.
Specially chosen $w_i$'s in (\ref{lws.equ}) will recover the famous LMS and the LTS in \cite{R84}, and LST in Zuo and Zuo \cite{ZZ23}.\vs
Previous weight functions in the literature are either constant ($0$ and $1$), or independent of $r_i(\boldsymbol{\beta})$, or do not down-weight large  residuals sufficiently, or non-differentiable.
 But there is much room for smooth wight functions such as those ones given in \cite{Z03}, \cite{Zetal04}, and \cite{ZC05}.  
To avoid drawbacks above, we propose using differentiable $w(r)$, which would assign weight one to $r_i$s 
which lies close to the center of data (all $r_i$s) cloud. The other points which lie on the outskirts of data (all $r_i$s) cloud could be viewed as outliers, so a lower positive weight should
be given.  This would balance efficiency with robustness. More discussions on $w$ and $\widehat{\bs{\beta}}_{wls}$ are carried out in section \ref{Sec.2}, where
an ad hoc choice of the weight function with the above property in mind will be introduced.
\vskip 3mm
The rest of this article is organized as follows. Section \ref{Sec.2} introduces 
a class of differentiable weight functions and a class of 
 weighted least squares estimators. Section \ref{Sec.3} establishes the existence of $\widehat{\bs{\beta}}_{wls}$ and  studies its properties including its finite sample breakdown robustness. Section \ref{Sec.4} discusses the computation of  $\widehat{\bs{\beta}}_{wls}$.
Section \ref{Sec.5} presents some concrete examples, comparing the performance of $\widehat{\bs{\beta}}_{wls}$ with other leading estimators. Section \ref{Sec.6} ends the article with some concluding remarks. Long proofs of the main results are deferred to an Appendix. 
\vs
\section{A class of  weighted least squares} \label{Sec.2}
\vs
\subsection{A class of weight functions}
An ad hoc choice of the weight function with the property mentioned in section 1
takes the form of
\vspace*{-5mm}
\bec
\begin{figure}[!ht]
    \centering
    \begin{subfigure}[!ht]{0.45\textwidth}
    \includegraphics[width=\textwidth]{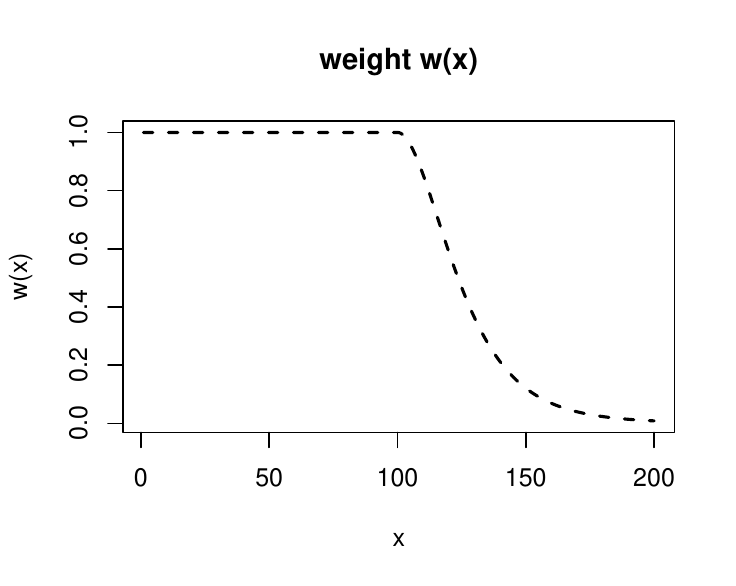}
     \end{subfigure}
     \hspace*{2mm}
    \begin{subfigure}[!ht]{0.45\textwidth}
    \vspace*{1mm}
    \includegraphics[width=\textwidth]{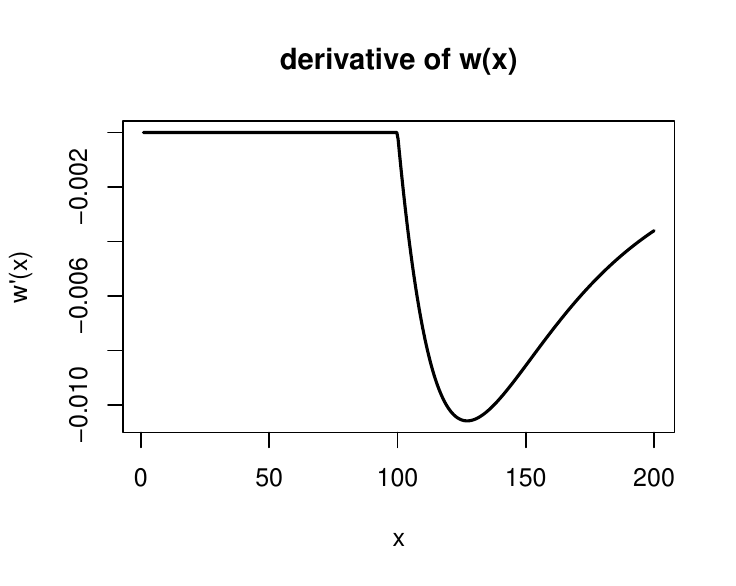}
   \end{subfigure}
   \caption{\footnotesize {Weight function $w(x)$ when $k=5$ and  $c=100$. ~Left: $w(x)$, right: $w'(x)$.}}
  \label{fig:w-w'}
\vspace*{-2mm}
\end{figure}
\enc

\be
w(x)=\mathds{1}(|x|\leq c)+\mathds{1}(|x|>c)\frac{e^{-k(1-c/|x|)^2}-e^{-k}}{1-e^{-k}},~~ \forall c, k>0, \label{weight.equ}
\ee
where tuning parameter $k>1$ is a positive number (say, between 1 and 10) controlling the steepness of the exponentially decreasing (see Figure \ref{fig:w-w'}), the larger the k, the steeper the curve; tuning parameter $c$ is  the point where the weight function will change from constant one to exponentially decreasing, $c>1$ usually can be set to be a large positive number (say $10$) or it can be residuals dependent, say $50\%$ or $75\%$ percentile of the residuals,  a larger $c$ is favorable for a higher efficiency.
\vs\vs
One of examples of $w(x)$ is given
in Figure \ref{fig:w-w'},  where $w(x)$ and its derivative are given and $k=5$ and $c=100$. For a general $w(x)$,
it is straightforward to verify that
\bi
\item[\tb{P1}] $w(x)$ is  twice differentiable and $0< w(x)\leq 1$. When $x\to \infty$, $w(x)$ is asymptotically equivalent to $\alpha (e^{\gamma x^{-1}}-1)$ for some positive constants $\alpha$ and $\gamma$.
\item[\tb{P2}] If $r_i \to \infty$, 
 then  $w(r^2_i/c^*) r^2_i \to 2ckc^*/(e^{k}-1)$, where $c^*=\mbox{Med}_i\{y^2_i\}$, the median of $\{y^2_1, y^2_2, \cdots, y^2_n\}$.
\ei

\subsection{Weighted least squares estimators}
With the weight function given above, we are ready to specify the 
weighted least squares estimator in (\ref{lws.equ}) with more details
\be
\widehat{\bs{\beta}}_{wls}=\arg\min_{\bs{\beta}\in \R^p} \sum_i^n w_ir^2_i(\bs{\beta}),\label{owls.equ}
\ee
where
weight $w_i:=w(r^2_i/c^*)$ with $w(x)$ being a weight function in (\ref{weight.equ}) and $c^*$ defined in \tb{P2}. 
\vs
The behavior of function $w(r^2/c^*)r^2$  when $r>c$ for different $c^*$s is illustrated in Figure (\ref{fig:w(r)r-sq}) below.
\vspace*{-1mm}
\bec
\begin{figure}[!ht]
    \centering
    \begin{subfigure}[!ht]{0.45\textwidth}
    \includegraphics[width=\textwidth]{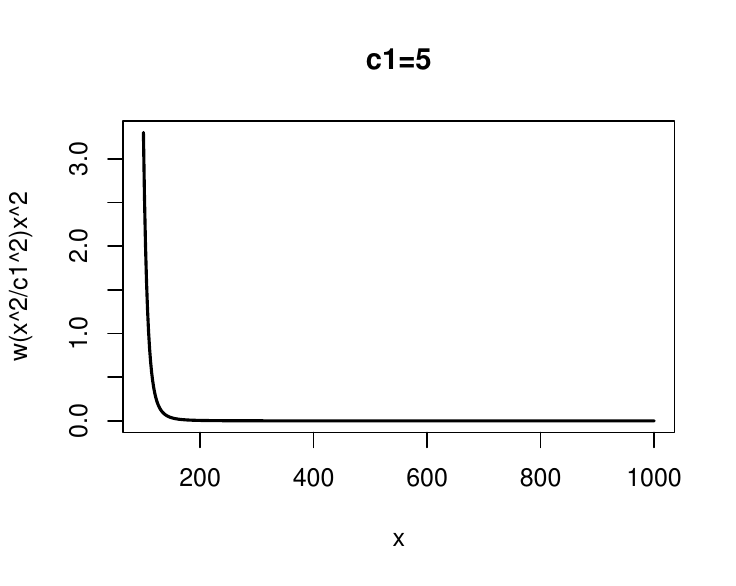}
     \end{subfigure}
     \hspace*{2mm}
    \begin{subfigure}[!ht]{0.45\textwidth}
    \vspace*{1mm}
    \includegraphics[width=\textwidth]{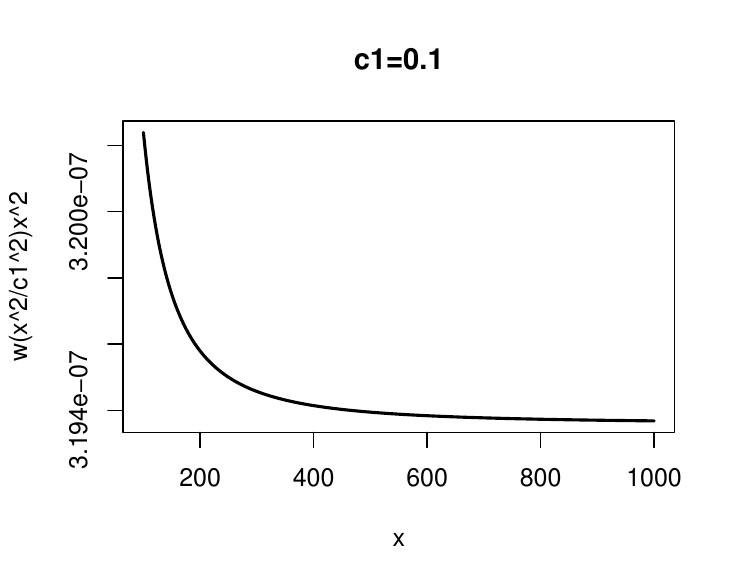}
   \end{subfigure}
   \caption{\footnotesize {Behavior of function $w(x^2/c^*)x^2$ when $k=5$ and $c=100$, $x>c$.}}
  \label{fig:w(r)r-sq}
\vspace*{-1mm}
\end{figure}
\enc

\vspace*{-8mm}
\section{Properties of the $\widehat{\bs{\beta}}_{wls}$} \label{Sec.3}
\subsection{Existence}
Does the minimizer of the objective function $O(\bs{\beta}, \bs{z}^{(n)}):=\sum_i^n w_ir^2_i(\bs{\beta})$ on the right-hand side (RHS) of (\ref{owls.equ}) exist? We now formally address this.
We need the following assumption:\vs 
\tb{A1:} For a given sample $\bs{z}^{(n)}:=\{(\bs{z}_i)_{i=1}^n\}=\{(\bs{x}^{\top}_i, y_i)^{\top}, i\in\{1,2,\cdots, n\}\}$ and any $\bs{\beta}\in \R^p$, all points $\{(\bs{x}^{\top}_i, y_i)^{\top}\}$ with
 $r_i$s satisfying $r^2_i/c^*\leq c$ do not lie in a vertical hyperplane.  
\vs
The assumption holds true with probability one if the sample comes from a distribution that has a density. Now we have the following
existence result.\vs
\vs
\noin
\tb{Theorem 3.1} If \tb{A1} holds true, then the minimizer $\widehat{\bs{\beta}}_{wls}$ of $O(\bs{\beta}, \bs{z}^{(n)})$ always exists. 
\vs
\noindent
\tb{Proof:} ~ see the Appendix. \hfill \pend


\vs
\subsection{equivariance}
\vs
Desirable fundamental properties of regression estimators include {regression, scale}, and {affine equivarince}. For $\mathbf{x} \in \R^{n\times (p-1)}$ and $\mathbf{y}\in \R^n$,
a regression estimator $\boldsymbol{\hat \beta}:=\mb{t}(\mathbf{\mb{w}, y})$ with $\mb{w}=(1,\bs{x}^{\top})^{\top}$ satisfying
\be
\mb{t}(\mathbf{\mb{w}, \mb{y}+\mb{b}^{\top}\mb{w}})=\mb{t}(\mathbf{\mb{w}, \mb{y}})+\mathbf{b},~ \forall \mathbf{b}\in \R^p;
\ee
\be
\mb{t}(\mathbf{w},s\mathbf{y})=s\mb{t}(\mb{w},\mb{y}),~ \forall s\in \R;
\ee
\be
\mb{t}(\mathbf{A^{\top}w, \mb{y}})=\mathbf{A^{(-1)}} \mb{t}\mathbf{(w,y)}, ~\forall ~ \mbox{nonsingular} \mathbf{A} \in \R^{p \times p}.
\ee
is called \emph{regression}, \emph{scale}, and \emph{affine} equivariant, respectively (see page 116 of \cite{RL87}). All aforementioned regression estimators are regression, scale, and affine equivariant.
\vs
\noin
\tb{Theorem 3.2}~ $\widehat{\bs{\beta}}_{wls}$ defined in (\ref{lws.equ}) is  \emph{regression}, \emph{scale}, and \emph{affine} equivariant.
\vs
\noindent
\tb{Proof:} Notice the identities: $r_i=y_i-\bs{w}^{\top}_i\bs{\beta}=y_i+\bs{b}^{\top}\bs{w}_i-\bs{w}^{\top}_i(\bs{\beta}+\bs{b})$, $sr_i=sy_i-\bs{w}^{\top}_i(s\bs{\beta})$, 
and $r_i=y_i-(\bs{A}^{\top}\bs{w}_i)^{\top}\bs{A}^{-1}\bs{\beta}$ meanwhile $r^2_i/c^*$ is regression, scale,
and affine invariant. The desired result follows. \hfill \pend
\vs
\subsection{Robustness}
As an alternative to the least-squares $\widehat{\bs{\beta}}_{ls}$, is the $\widehat{\bs{\beta}}_{wls}$ more robust?\vs  The most prevailing quantitative measure of global robustness of any location or regression estimators in the finite sample practice is the \emph{finite sample breakdown point} (FSBP), introduced by Huber and Donoho (1983) \cite{DH83}. 
\medskip

Roughly speaking, the FSBP is the minimum fraction of `bad' (or contaminated) data that the estimator can be affected to an arbitrarily large extent. For example, in the context of estimating the center of  a data set,
the sample mean has a breakdown point of $1/n$ (or $0\%$), because even one bad observation can change the mean
by an arbitrary amount; in contrast, the sample median has a breakdown point of $\lfloor(n+1)/2\rfloor/n$ (or $50\%$), where $\lfloor \cdot\rfloor$ is the floor function.
\vs

\noindent
\textbf{Definition 3.1} \cite{DH83} ~
The finite sample \emph{replacement breakdown point} (RBP) of a regression estimator $\mb{t}$ at the given sample
$\mb{z}^{(n)}=\{z_1,\cdots,z_n\}$, where $\bs{z}_i:=(\bs{x}^{\top}_i, y_i)^{\top}$, is defined  as
\begin{equation}
\text{RBP}(\mb{t},\mb{z}^{(n)}) = \min_{1\le m\le n}\bigg\{\frac{m}{n}: \sup_{\mb{z}_m^{(n)}}\|\mb{t}(\mb{z}_m^{(n)})- \mb{t}(\mb{z}^{(n)})\| =\infty\bigg\},
\end{equation}
where $\mb{z}_m^{(n)}$
denotes an arbitrary contaminated sample by replacing $m$ original sample points in $\mb{z}^{(n)}$ with arbitrary points in $\R^{p}$.
 Namely, the RBP of an estimator is the minimum replacement fraction that could drive
the estimator beyond any bound.  It turns out that both $L_1$ (least absolute deviations) and $L_2$ (least squares) estimators have RBP $1/n$ (or $0\%$), the lowest possible value
 whereas $\widehat{\bs{\beta}}_{wls}$ can have $( \lfloor( n - p ) / 2 \rfloor + 1)/n$ (or $50\%$), the highest possible value for any regression equivariant estimators (see p. 125 of \cite{RL87}).
\vskip 3mm
We shall say  $\mb{z}^{(n)}$  is\emph{ in general position}
when any $p$ of observations in $\mb{z}^{(n)}$ gives a unique determination of $\bs{\beta}$.
In other words, any (p-1) dimensional subspace of the space $(\bs{x^{\top}}, y)^{\top}$ contains at most p observations of
$\mb{z}^{(n)}$.
When the observations come from continuous distributions, the event ($\mb{z}^{(n)}$ being in general position) happens with probability one.\vs
\noin
\tb{Theorem 3.3} Assume  that \tb{A1} holds true,  $n>p$, and
 $\mb{z}^{(n)}$ in general position. Then 
\be
\text{RBP}(\widehat{\bs{\beta}}^n_{wls}, \mb{z}^{(n)})=\left\{
\begin{array}{ll}
\lfloor (n+1)/2\rfloor\big/n, & \text{if $p=1$,}\\[1ex]
(\lfloor{(n-p)}/{2}\rfloor+1)\big/n,& \text{if $p>1$.}\\
\end{array}
\right. \label{T*-bp.eqn}
\ee
\vs
\noin
\tb{Proof:} ~ see the Appendix. \hfill \pend
\vs
\noin
We need the following important result for the proof of Theorem 3.3.

\vs
\noin
\tb{Lemma 3.1} For any $r^2_i>r^2_j>c^*c$, $w(r^2_i/c^*)r^2_i<w(r^2_j/c^*)r^2_j$ when $r^2_j\to \infty$.
\vs
\noin
\tb{Proof:} ~ see the Appendix. \hfill \pend

\vs\vs
\noin
\tb{Remarks 3.1} 
The RBP result in Theorem 3.3 is the highest possible breakdown point for any regression equivariant estimators in the literature (see p. 125  of \cite{RL87}).  There are very few regression estimators that possess the highest breakdown point robustness.
\hfill  \pend

\section{Computation of the WLS} \label{Sec.4}
Now we address the most important issue with high breakdown point estimator, the computation of the estimator. The objective function in (\ref{ls.equ}) is
\be O(\bs{\beta}):=O(\bs{\beta}, \bs{z}^{(n)})=\sum_{i=1}^n w(r^2_i/c^*)r^2_i,\label{objective.equ}\ee
which is differentiable with respect to $\bs{\beta}$ since weight function $w(x^2/c^*)$ is twice differentiable with
\begin{align}
w'(x)&=\alpha^* e^{-k(1-c/|x|)^2}(1-c/|x|)\mbox{sgn}(x)/x^2\mathds{1}(|x|>c), \nonumber\\[2ex]
 w''(x)&={\alpha^*}e^{-k(1-c/|x|)^2}\big({-2kc}(1-c/|x|)^2/|x|-(2-3c/|x|)\big)/x^3\mathds{1}(|x|>c),
\end{align}
where $\alpha^*=-2kc/(1-e^{-k})$. \vs
The problem in (\ref{lws.equ}) belongs to an unconstrained minimization. This type of problem has been thoroughly discussed  and studied in the literature. Common approaches to find the solution include (i) methods utilizing  first-order derivatives(gradient descent/steepest descent/conjugate gradient)  (ii) methods using second-order derivatives (Hessian matrix) (Newton's method) (iii) Quasi-Newton method, see \cite{BV04} and \cite{EHL01}. We will select the conjugate gradient  for speed/efficiency and accuracy consideration. \vs

Note that
\begin{align}
\nabla O(\bs{\beta})=\frac {\partial O(\bs{\beta}) }{\partial\bs{\beta} } 
&=\sum_{i=1}^n(w'(r^2_i/c^*)r^2_i+c^*w(r^2_i/c^*))\frac{\partial r^2_i/c^*}{\partial \bs{\beta}} \nonumber \\[2ex]
&=\sum_{i=1}^n(w'(r^2_i/c^*)r^2_i+c^*w(r^2_i/c^*))2r_i/c^* (-\bs{w}^{\top}_i)\nonumber \\[1ex]
&=\sum_{i=1}^n -2r_i/c^*(w'(r^2_i/c^*)r^2_i+c^*w(r^2_i/c^*))\bs{w^{\top}_i}.
\end{align}
\begin{align}
\nabla^2 O(\bs{\beta})=\frac {\partial^2 O(\bs{\beta}) }{\partial^2\bs{\beta} }
&=\frac{-2}{c^*}\sum_{i=1}^n\bs{w}^{\top}_i \frac{\partial r_i(w'(r^2_i/c^*)r^2_i+c^*w(r^2_i/c^*)}{\partial \bs{\beta}} \nonumber \\[1ex]
&=\frac{-2}{c^*}\sum_{i=1}^n\bs{w}^{\top}_i\bs{w}_i\Big(5\frac{r^2_i}{c^*}w'(\frac{r^2_i}{c^*})+w(\frac{r^2_i}{c^*})+2(\frac{r^2_i}{c^*})^2w''(\frac{r^2_i}{c^*})\Big)\nonumber \\[1ex]
&=\bs{X}^{\top}_n\bs{W}\bs{X}_n,
\end{align}
where $\bs{X}^{\top}_n=(\bs{w}^{\top}_1, \cdots, \bs{w}^{\top}_n)$, $\bs{W}$ is a diagonal matrix with its ith diagonal entry $-2\gamma_i/c^*$ and $$\gamma_i=5\frac{r^2_i}{c^*}w'(\frac{r^2_i}{c^*})+w(\frac{r^2_i}{c^*})+2(\frac{r^2_i}{c^*})^2w''(\frac{r^2_i}{c^*}).$$
\vs
\noin
\tb{The algorithm for the conjugate gradient method (CGM)} is as follows: \vs
{\small 
\bi
\item[(i)] \tb{Step 1}. Pick a $\bs{\beta}^0$ (which can be an LS estimator, but for robustness the LTS (\cite{R84}) or LST (\cite{ZZ23}) is a better choice).
Set $\bs{v}^0=-\nabla O(\bs{\beta}^0)$. Set a tolerance $\varepsilon$. if $(\|v^0\|<\varepsilon)$ \{return $\beta^0$\}.
\item[(ii)] \tb{Step 2}. For $k=0, 1,\cdots, n-1$,
\bi
\item[a)] set $\bs{\beta}^{k+1}=\bs{\beta}^{k}+\alpha^{k}\bs{v}^{k}$, where $\alpha^k$ is the minimizer of $O(\bs{\beta}^k+\alpha\bs{v}^k)$ with respect to $\alpha$ (using backtracking line search, see page 464 of \cite{BV04}), or set
\[\alpha^k=- \nabla^{\top}O(\bs{\beta}^k)\bs{v}^k\big/(\bs{v}^k)^{\top}H(\bs{\beta}^k)\bs{v}^k, \]
where $H(\bs{\beta}^k)=\nabla^2(O(\bs{\beta}^k))$.
\item[b)]compute $ \nabla O(\bs{\beta}^{k+1})$, if $(\|\nabla O(\bs{\beta}^{k+1})\|<\varepsilon)$ \{return $\bs{\beta}^{k+1}$\}.
\item[c)] if ($k=n-1$) \{break\}; else set $v^{k+1}=-\nabla O(\bs{\beta}^{k+1})+\alpha^kv^k$, where
$$\alpha^k=\nabla^{\top}O(\bs{\beta}^{k+1})\nabla O(\bs{\beta}^{k+1})/\nabla^{\top}O(\bs{\beta}^{k})\nabla O(\bs{\beta}^{k})
$$
\ei
end for loop.
\item [(iii)] \tb{Step 3}. Replace $\bs{\beta}^0$ by $\bs{\beta}^n$ and go to step 1.
\ei
\section{Examples and Comparison}\label{Sec.5}

Now we investigate the performance of our new procedure WLS and compare it with some leading competitors including robust benchmark,
 the least trimmed squares LTS estimator, Rouseeuw \cite{R84} 
and the classical least squares LS estimator via some concrete examples.\vs  
\subsection{Performance criteria}

\noin
\tb{Empirical mean squared error (EMSE)}~~
For a general regression estimator $\mb{t}$,  We calculate
$\mbox{EMSE}:=\sum_{i=1}^R \|\mb{t}_i - \bs{\beta}_0\|^2/R$, the empirical mean squared error (EMSE) for $\mb{t}$. 
if $\mb{t}$ 
is  regression equivariant, then we can assume (w.l.o.g.) that the true parameter $\bs{\beta}_0=\mb{0}\in \R^p$ (see p.124 of \cite{RL87}). 
Here $\mb{t}_i$ is the realization of $\mb{t}$ obtained from the ith sample with size $n$ and dimension $p$, and replication number $R$ is usually set to be $1000$. 
\vs
\noin
\tb{Total time consumed for all replications in the simulation (TT)}~~ This criterion measures the speed of a procedure, the faster meanwhile accurate the better. One possible issue is the fairness of comparison of different procedures because different programming languages (e.g., C, Rcpp, Fortran, or R) are employed by different procedures. 


 \vs

\noin
\tb{Finite sample relative efficiency (FSRE)} ~~In the following we investigate via  simulation studies the finite-sample relative efficiency of the different robust alternatives of the LS w.r.t. the benchmark, the classical least squares line/hyperplane. The latter is optimal with normal models by the Gauss–Markov
theorem. We generate $R = 1,000$ samples from the linear regression model:
 $y_i = \beta_0 + \beta_1 x_i + \beta_{p-1}x_{p-1}+ e_i, i \in\{ 1, \cdots, n\}$
with different sample size $n$s and dimension $p$s, 
where $e_i\sim N(0, \sigma^2)$. The finite sample RE of a procedure is the percentage of EMSE of the LS divided by the EMSE of the procedure. \vs
All R code (downloadable via https://github.com/left-github-4-codes/WLS) for simulation, examples, and figures in this article  were run on a desktop Intel(R)Core(TM)
21 i7-2600 CPU @ 3.40 GHz. \vs

\vs
\bec
\vspace*{-5mm}
\begin{figure}[!ht]
	\includegraphics 
	[width=0.8\textwidth] %
	{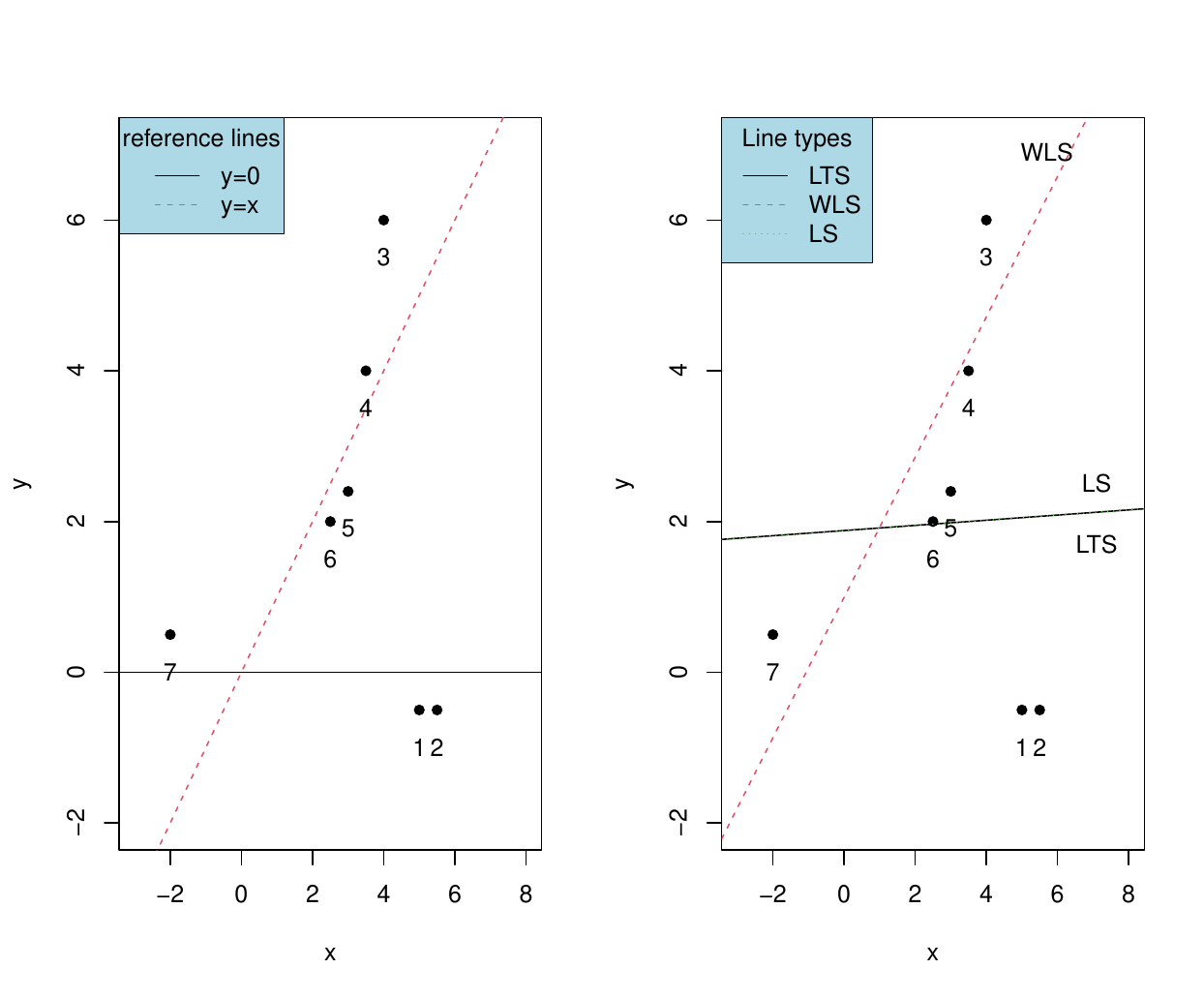}
	\caption{\footnotesize Left panel: plot of seven artificial points and two reference lines $y=0$ and $y=x$.  Right panel:  the same seven points are fitted by LTS, WLS, and the LS (benchmark). A solid black  line is LTS given by ltsReg. Red dashed line is given by WLS, and green dotted line is given by the LS - identical to LTS line in this case.}
	\label{fig-7-points}
	\vspace*{-3mm}
\end{figure}
\enc
\vs

\noin
\subsection{Examples}
\vs
\noin
\tb{Example 1} (\tb{Simple linear regression})   To take the advantage of graphical illustration of data sets and plots, we start with $p=2$, the simple linear regression case.
\vs
\noin
We borrow an artificial 7-point data set in Example 1.1 of  Zuo and Zuo \cite{ZZ23}. 
It is plotted in the left panel of the (a)
of Figure \ref{fig-7-points}. Two reference regression lines  ($y=0$) and  ($y=x$) are also provided. 
\vs
Inspecting the left panel of the Figure immediately reveals that points 1 and 2 seem to be outliers and the overall pattern of the data set is linear $y=cx$ with $c>0$. The right panel further reveals that both the LS and the LTS
are very sensitive to the outlying points whereas WLS still can catch the overall line pattern with the influence of two outliers.\vs

One might immediately argue that the example above has at least two drawbacks (i) data set is too small and (ii) it is purely artificial. In Figure \ref{fig-80}, the sample size is increased to $80$, $80$ highly correlated normal points with $30\%$ of them contaminated by other normal points.  Inspecting the Figure reveals that the 
three procedures capture the linear pattern perfectly in the left panel of the Figure for perfect bivariate normal points, while in the right panel, both
LTS and LS lines are drastically changed due to the 24 contaminating points.
\bec
\vspace*{-5mm}
\begin{figure}[!ht]
	\includegraphics 
	[width=0.8\textwidth] %
	{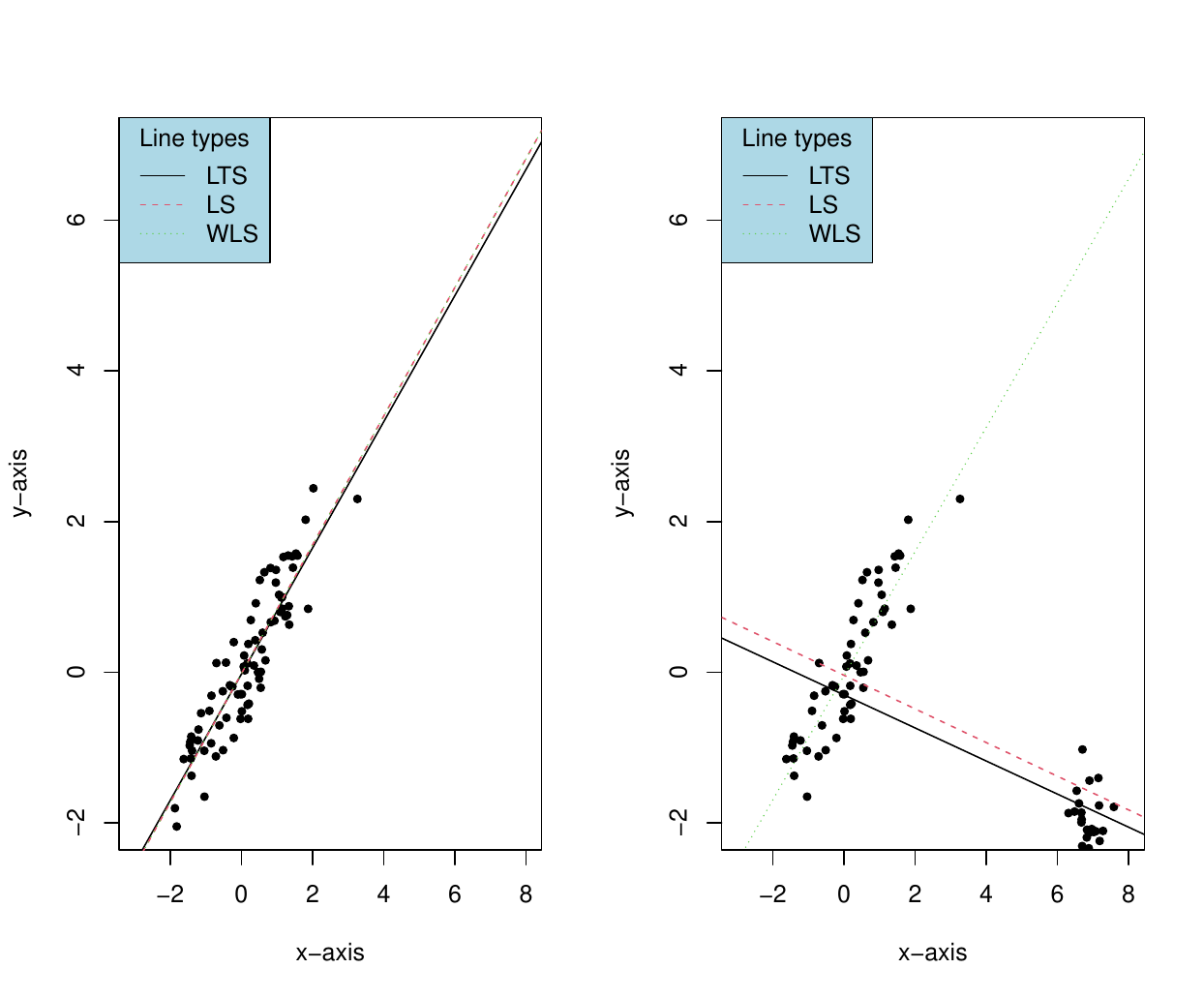}
	\caption{\footnotesize $80$ highly correlated normal points with $30\%$ of them  contaminated by other normal points. Left: scatterplot of the un-contaminated perfect normal data set and three almost identical lines. Right: LTS,  WLS, and LS lines for the contaminated data set. Solid black is LTS line, dotted green is the WLS, and dashed red is given by the LS - parallel to LTS line in this case.}
	\label{fig-80}
	\vspace*{-3mm}
\end{figure}
\enc

In practice, there are more cases with more than one independent variable, in the following we consider the case $p>2$.
\vs
\noin
\tb{Example 2} (\tb{Multiple linear regression with contaminated normal points})
Now we do not have the visual  advantage like in the $p=2$ case. To compare the performance of different procedures,
we have to appeal the performance measures discussed in section 5.1. \vs

We consider the contaminated highly correlated normal data points scheme. We generate $1000$ samples $\{\mb{z}_i=(\bs{x^{\top}_i},y_i)^{\top}, i\in \{1, \cdots, n\}\}$ with various $n$s  from the  normal distribution ${N}(\bs{\mu}, \bs{\Sigma})$,
where $\bs{\mu}$ is a zero-vector in $\R^p$, and $\bs{\Sigma}$ is a $p$ by $p$ matrix with diagonal entries being $1$ and off-diagonal entries being $0.9$. Then $\varepsilon\%$ of them are contaminated by $m=\lceil n\varepsilon\rceil$ points, where $\lceil\cdot\rceil$ is the ceiling function. We randomly select $m$ points of $\{\bs{z}_i$, $i\in\{1,\cdots, n\}\}$ and replace them by $(3,3, \cdots, 3, -3)^{\top}$. 
\vs
The performance of the CGM in Section \ref{Sec.4} (or rather any iterative procedure)  severely depends on the initial given point $\bs{\beta}^0$.
In light of its cyclic feature of the CGM for nonquadratic objective function (see page 195 of \cite{EHL01})
and our extensive empirical simulation experience,
the performance of the $\bs{\beta}$ return by the CGM usually is no much different (or better) than that of the initially selected $\bs{\beta}^0$.
To achieve better performance for the WLS, we modified the LST of Zuo and Zuo \cite{ZZ23} and utilized it as the initial (and  final) $\bs{\beta}$ for CGM. Results for the three methods and different $n$s and $p$s and contamination levels $\varepsilon$ are listed in Table \ref{tab.one}.\vs

Inspecting the Table reveals that (i) LS is the best performer for pure normal data sets and the fastest in all cases considered,
it, however, becomes the worst performer when there is contamination (except the $30\%$ contamination case where the LTS surprisingly becomes the worst performer, in theory both WLS and LTS can resist up to $50\%$ contamination without breakdown). (ii) LTS is well-known for its robustness and computation speed, however, it has no advantage in these two aspects over WLS in all cases considered. (iii) For pure normal data ($\varepsilon=0\%$), WLS is almost as efficient as the LS whereas LTS has RE about $70-80\%$. For contaminated data, WLS is the most efficient performer followed by LTS as the second and LS as the worst one (except when there are $30\%$ contaminated, then LS performs even better than LTS in terms of RE).

\vs
\begin{table}[!h]
	\centering
	~~ Normal data sets, each with $\varepsilon $ contamination rate\\
	\bec
	\begin{tabular}{c c c c c cccc }
		
		~p~ & ~n~&method&EMSE&TT&RE&~~~~EMSE&TT&RE \\
		\hline\\[0.ex]
		~ & ~&~&~$\varepsilon=0\%$&~&~&~~~~$\varepsilon=10\%$~&~&\\[1ex]
		& 50& lts &0.4025&16.934& 0.8106&~~~0.3751&16.191&2.7536\\
	5	& 50& wls &0.3272&7.3039& 0.9972&~~~0.3382&8.7148&3.0546\\
		& 50& ls  &0.3263&1.4518& 1.0000&~~~1.0330&1.3750& 1.0000\\[2ex]
			~ & ~&~&~$\varepsilon=20\%$&~&~&~~~~$\varepsilon=30\%$~&~&\\[1ex]
		& 50& lts &0.5809&16.358&3.3078&~~~28.103&17.767&0.0966\\
	5	& 50& wls &0.3518&11.191&5.4619&~~~0.3687&14.840&7.3613\\
		& 50& ls  &1.9216&1.5301&1.0000&~~~2.7140&1.3143&1.0000\\[2ex]
	~ & ~&~&~$\varepsilon=0\%$&~&~&~~~~$\varepsilon=10\%$~&~&\\[1ex]	
      & 100& lst &0.2979&49.193&0.7119&~~~0.2656&50.377&5.0190\\
 10   & 100& wls &0.2122&10.856&0.9993&~~~0.2254&14.025&5.9139\\
      & 100& ls  &0.2121&1.2969&1.0000&~~~1.3330&1.3162&1.0000\\[2ex]
~ & ~&~&~$\varepsilon=20\%$&~&~&~~~~$\varepsilon=30\%$~&~&\\[1ex]
     & 100& lts &0.3015&55.077&8.3664&~~~41.552&64.572&0.0834\\
10   & 100& wls &0.2420&19.484&10.424&~~~0.2607&26.087&13.297\\
     & 100& ls  &2.5228&1.2969&1.0000&~~~3.4673&1.3125&1.0000\\[2ex]
 	~ & ~&~&~$\varepsilon=0\%$&~&~&~~~~$\varepsilon=10\%$~&~&\\ [1ex]
 	
	& 200& lts &0.2186&261.08&0.7347&~~~0.2020&294.67&7.2605\\
20  & 200& wls &0.1608&27.493&0.9983&~~~0.1734&36.987&8.4575\\
    & 200& ls  &0.1606&1.4304&1.0000&~~~1.4669&1.4703&1.0000\\[2ex]
~ & ~&~&~$\varepsilon=20\%$&~&~&~~~~$\varepsilon=30\%$~&~&\\[1ex]
    & 200& lts  &0.2022&392.62&13.611&~~~32.241&844.60&0.1208\\
 20 & 200& wls  &0.1915&49.720&14.372&~~~0.2134&66.281&18.247\\
    & 200& ls   &2.7524&1.4693&1.0000&~~~3.8936&1.4281& 1.0000\\				
		\hline
	\end{tabular}
	\enc
	\caption{\footnotesize EMSE, TT (seconds), and  RE for LTS, WLS, and LS based on all $1000$ samples for various $n$s, $p$s, and contamination levels.}
	\label{tab.one}
\end{table}
\vs

\vspace*{-5mm}
\noin
\tb{Example 3} (\tb{Performance when $\bs{\beta}^0$ is given}). In the calculation of EMSE above, one assumes that
$\bs{\beta}^0=\bs{0}$ in light of regression equivariance of an estimator $\bs{t}$. In this example, we will
provide $\bs{\beta}^0$ (for convenience write it as $\bs{\beta}_0$) and calculate $y_i$ using the formula $y_i= (1,\bs{x}^{\top}_i)\bs{\beta}^{\top}_0+e_i$, where we simulate
$\bs{x}_i$ from a normal distribution with zero mean vector and identical covariance matrix. $e_i$ follows a standard normal
 distribution. 

\bec
\vspace*{-8mm}
\begin{figure}[!ht]
	\includegraphics [width=14cm, height=10cm]%
	{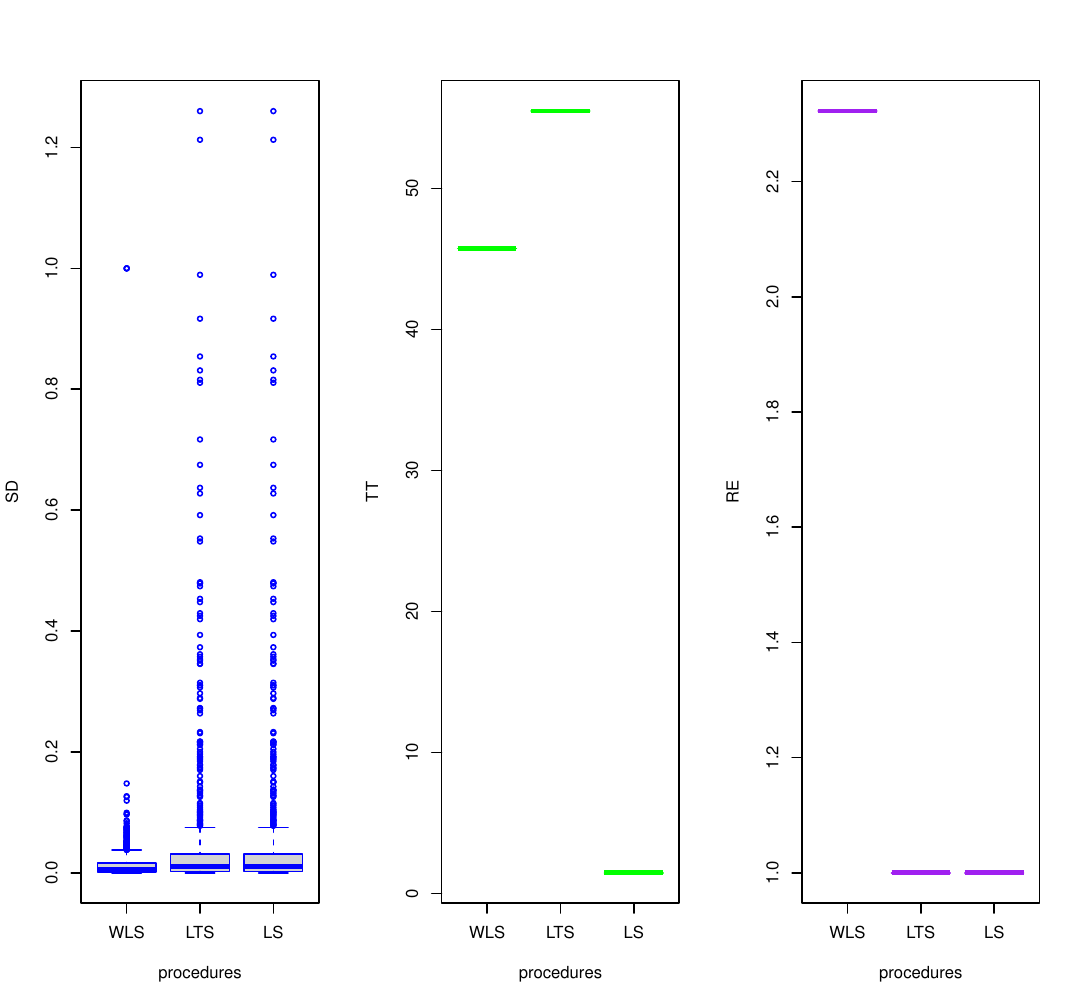}
	\caption{\footnotesize Performance of three procedures with respective $1000$ normal samples (points are highly correlated) with $p=10$ and $n=100$, each sample suffers $10\%$ contamination.}
	\label{boxplot}
	\vspace*{-5mm}
\end{figure}
\enc

 We set $p=10$, $n=100$ and $\bs{\beta}_0=(1,1,1,1,1,-1,-1,-1,-1,-1)$. There is a $10\%$ contamination for each of $1000$ normal samples (generated as in Example 2) with  the contamination scheme as: we randomly select $m=\lceil n\varepsilon\rceil$ points out of $\{\bs{z}_i$, $i\in\{1,\cdots, n\}\}$ and replace them by $(3.5,3.5, \cdots, 3.5)^{\top}$.
We then calculate the squared deviation (SD) $(\widehat{\bs{\beta}}_i-\bs{\beta}_0)^2$ for each sample, the total time (TT) consumed by each procedures for all $1000$ samples, and the relative efficiency (RE) (the ratio of EMSE of LS vs EMSE of a procedure). The  performance  three procedures for different  criteria are displayed via boxplot in Figure \ref{boxplot}.
\vs

Inspecting the Figure reveals that (i) in terms of squared deviations, LTS and LS perform almost the same, both have a wide spread
and a high EMSE whereas WLS has a much smaller EMSE (in fact, the EMSE for the three are $(1.049303, 2.568752, 2.568752)$).
(ii) in terms of total time consumed, LS is the absolute winner, and LTS is the loser, WLS is  much better than LTS. (iii) in terms of relative efficiency, LTS is the loser (performs as bad as the LS) whereas WLS earns the trophy and is much more robust against $10\%$ contamination.\vs

Up to this point, we have dealt with synthetic data sets. Next, we investigate the performance of WLS, LTS,  and LS with respect to real data sets in high dimension.
\vs

\noin
\tb{Example 4}  (\tb{Performance for a large real data set}) Boston housing is a famous data set (see \cite{HR87}) and studied by many authors with different emphasizes (transformation, quantile, nonparametric regression, etc.) in the literature. For a more detailed description of the data set, see http://lib.stat.cmu.edu/datasets/.\vs
The analysis
reported here did not include any of the previous results, but consisted of just a straight linear regression of the dependent variable (median price of a house) on the thirteen explanatory variables as might be used in an initial exploratory analysis of a new data set.
We have sample size $n=506$ and dimension $p=14$.
\vs
We assess the performance of the LST,  the WLS, and LS  as follows. 
Since some methods depend on randomness, so we run the computation  $R=1000$ times to alleviate the randomness.
(i) We  compute the $\widehat{\bs{\beta}}$ with different methods, we do this $1000$ times. 
(ii) We calculate the total time consumed (in seconds) by different methods for all replications, and the EMSE (with true $\bs{\beta}_0$ being replaced by the sample mean of $1000$ $\widehat{\bs{\beta}}$s from (i)), which is the sample variance of all $\widehat{\bs{\beta}}$s up to a factor $1000/999$.
The results are reported in Table \ref{tab.two}. 
\vs

\vs
\begin{table}[!ht]
	\centering
	\bec
	\begin{tabular}{c c c c   }
		
		~performance measure~ & LTS~~&~WLS~& ~LS~  \\
		\hline\\[0.ex]
		EMSE&  41.543  &0.0000& 0.0000 \\
	   TT&157.04 &157.04&1.6110\\
		RE&   0 &NaN& NaN \\
		\hline
	\end{tabular}
	\enc
	\caption{\footnotesize EMSE, TT (seconds), and  RE for LTS,  WLS, and LS based on Boston housing  real data set.}
	\label{tab.two}
\end{table}

Inspecting the table reveals that (i)  WLS and LS produce the same $\widehat{\bs{\beta}}$ for each sample, so there is no
variance whereas this is not the case for LTS. (ii) LS is the fastest runner followed by LTS and WLS.
(iii) the relative efficiency of LTS is $0\%$ since the sample variance of LS is $0$ whereas the RE of  WLS and LS is undefined
(not a number), since $0$ appeared in the denominator. On the other hand, one can interpret WLS as good as LS in this case $100\%$.


\section{Concluding remarks}\label{Sec.6}
With a novel weighting scheme, the proposed weighted least squares estimator performs as efficient as the classic least squares (LS) estimator
for perfect normal data and much more efficient than the robust least trimmed squares (LTS) estimator; whereas it is much more robust than the LS when there are contamination or outliers, it performs as robust as the LTS  while much more efficient than LTS when there are outliers. It could serve as a robust alternative to the least squares estimator in practice.
\vs
\vs

\begin{center}
	{\textbf{\large Acknowledgments}}
\end{center}
The authors thank Prof. Wei Shao 
for 
insightful comments and stimulus discussions.

\vs
\vs

\bec
\tb{\large Appendix}
\enc

\vs\vs

\noin
\tb{Proof of Theorem 3.1}
For a given  $\bs{z}^{(n)}$ and any $\bs{\beta}$, 
write $M: =\sum_{i=1}^n y^2_i\geq \sum_{i=1}^n w(y_i^2/c^*) y^2_i=O( \bs{0}_{p \times 1}, \bs{z}^{(n)})$.
For a given $\bs{\beta}\in\R^p$, hereafter assume that $H_{\bs{\beta}}$ is the hyperplane determined by $y=\bs{w}^{\top}\bs{\beta}$ and let $H_h$ be the horizontal hyperplane (i.e., $y=0$, the $\bs{w}$-space).
\vs
Partition the space of $\bs{\beta}$s into two parts: $S_1$ and $S_2$, with $S_1$ containing all $\bs{\beta}$s such that $H_{\bs{\beta}}$ and $H_h$ are parallel and $S_2$ consisting of the rest of  $\bs{\beta}$s so that $H_{\bs{\beta}}$ and $H_h$ are not parallel.
\vs
If one can show that there are minimizers of $O(\bs{\beta}, \bs{z}^{(n)})$ over $S_i$ $i=1,2$, respectively, then one can have an overall minimizer.
\vs
Over $S_1$, $\bs{\beta}=(\beta_0, \mb{0}^{\top}_{(p-1) \times 1})^{\top}$ and $r_i=y_i-\beta_0$. If the minimizer does not exist, then it means that any bounded $\beta_0$ can not minimize  $O(\bs{\beta}, \bs{z}^{(n)})$, the absolute value of the minimizer
$\beta^*_0$ must be greater than any $M^*>0$. We seek a contradiction now. 
Denote the minimizer as ${\bs{\beta}}^*=(\beta^*_0, \mb{0}^{\top}_{(p-1)\times 1})^{\top}$.  Define ${\bs{\beta}}^*_1=(2\beta^*_0, \mb{0}^{\top}_{(p-1)\times 1})^{\top}$
then it is readily seen that $r^2_i({\bs{\beta}}^*_1)> r^2_i({\bs{\beta}}^*)$ for large enough $\beta^*_0$. By lemma 3.1 below,
one has $O({\bs{\beta}}^*, \bs{z}^{(n)})>O({\bs{\beta}}^*_1, \bs{z}^{(n)})$. A contradiction
has researched.
\vs

Over $S_2$, denote by $l_{\bs{\beta}}$  the
intersection part of $H_{\bs{\beta}}$ with the horizontal hyperplane $H_h$ 
(we call it a hyperline, though it is $p-1$-dimensional).
Let $\theta_{\bs{\beta}}\in(-\pi/2, \pi/2)$ be the acute angle between the $H_{\bs{\beta}}$ and $H_h$ (and $\theta_{\bs{\beta}}\not =0$).
Consider two cases. 
\vs
\tb{Case I}. All 
$\bs{w}_i=(1, \bs{x}^{\top}_i)^{\top}$ with $r^2_i/c^*\leq c$  on the hyperline $l_{\bs{\beta}}$, where $r_i=y_i-\bs{w}^{\top}_i\bs{\beta}$. Then we have a vertical hyperplane that is perpendicular to the horizontal hyperplane $H_h$ $(y=0)$ and intersect $H_h$ at
$l_{\bs{\beta}}$,  
But this contradicts \tb{A1}. We only need to consider the other case.
\vs
\tb{Case II}. Otherwise, define
$$
\delta=\frac{1}2\inf\{\tau, \mbox{such that $N(l_{\bs{\beta}}, \tau)$ contains all $\bs{w}_i$ with $r^2_i/c^*\leq c$}
\},
$$
where $N(l_{\bs{\beta}}, \tau)$ is the set of points in $\bs{w}$-space such that each distance to the $l_{\bs{\beta}}$ is no greater than $\tau$. Clearly, $0<\delta<\infty$ (since $\delta=0$ has been covered in \tb{Case I} and $\delta\leq \max_i\{\|\bs{w}_i\|\}<\infty$ where $i$ satisfying $r^2_i/c^*\leq c$ ,  the first inequality follows from the fact that hypotenuse is always longer than any legs).
\bec
\vspace*{-5mm}
\begin{figure}[!ht]
	\includegraphics
	[scale=0.60]
	{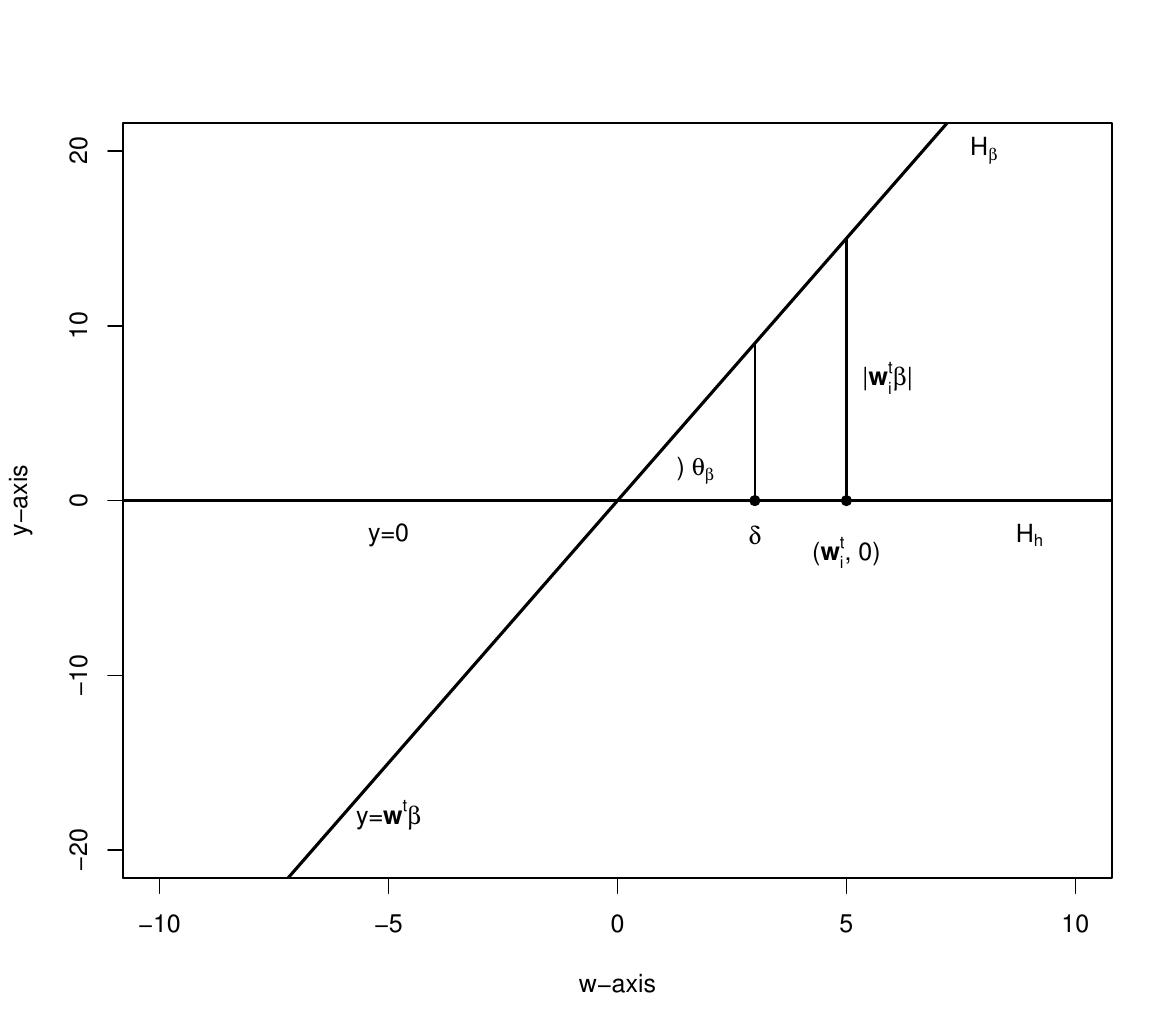}
	\caption{\footnotesize A two-dimensional vertical cross-section (that goes through points $(\bs{w}^t_i, 0)$ and $(\bs{w}^t_i, \bs{w}_i^t\bs{\beta})$) of a figure in $\R^p$ ($\bs{w}_i^t=\bs{w}^{\top}_i$).
		Hyperplanes $H_h$ and $H_{\bs{\beta}}$ intersect at hyperline $l_{\bs{\beta}}$ (which does not necessarily pass through $(\bs{0},0)$, here just for illustration).
		The vertical distance from point $(\bs{w}^t_i, 0)$ to the hyperplane $H_{\bs{\beta}}$, $|\bs{w}^t_i\bs{\beta}|$, is greater than $\delta|\tan(\theta_{\bs{\beta}})|$. }
	\label{fig-4-proof}
	\vspace*{-2mm}
\end{figure}
\enc

We now show that when $\|\bs{\beta}\|> (1+\eta)\sqrt{M}/\delta$, where $\eta>1$ is a fixed number, then
\be
O(\bs{\beta}, \bs{z}^{(n)})=\sum_{i=1}^n w(r^2_i/c^*) r^2_i(\bs{\beta})> M \geq  O( \bs{0}_{p \times 1}, \bs{z}^{(n)}).\label{proof.inequality}
\ee
That is, for the solution of minimization of (\ref{owls.equ}), one only needs to search over the ball $\|\bs{\beta}\|\leq(1+\eta)\sqrt{M}/\delta$, a compact set.
Note that $O(\bs{\beta}, \bs{z}^{(n)})$ is continuous in
$\bs{\beta}$ since $r_i(\bs{\beta})$ and $w(r^2_i/c^*)$ are. 
Then the minimization problem certainly has a solution over the compact set.
\vs
The proof is complete if we can show (\ref{proof.inequality}) when $\|\bs{\beta}\|> (1+\eta)\sqrt{M}/\delta$. It is not difficult to see that there is at least one $i_0$ such that $r^2_{i_0}/c^*\leq c$ and
$\bs{w}_{i_0} \not \in N(l_{\bs{\beta}}, \delta)$ since otherwise it contradicts the definition of $\delta$ above.  Note that $\theta_{\bs{\beta}}$ is the angle between the normal vectors $(-\bs{\beta}^{\top}, 1)^{\top}$ and $(\bs{0}^{\top}, 1)^{\top}$ of hyperplanes $H_{\bs{\beta}}$ and $H_h$, respectively. Then  $|\tan{\theta_{\bs{\beta}}}|=\|\bs{\beta}\|$ (see \cite{Z23}) and (see Figure \ref{fig-4-proof}) $$|\bs{w}^{\top}_{i_0}\bs{\beta}|>\delta|\tan{\theta_{\bs{\beta}}}|=\delta\|\bs{\beta}\|>(1+\eta)\sqrt{M}.$$
Now we have
\begin{align}
|r_{i_0}(\bs{\beta})|=|\bs{w}^{\top}_{i_0}\bs{\beta}-y_{i_0}| 
&\geq\big||\bs{w}^{\top}_{i_0}\bs{\beta}|-|y_{i_0}|\big| 
> (1+\eta)\sqrt{M}-|y_{i_0}|.
\end{align}
\vspace*{-5mm}
Therefore, \vs
\begin{align}
O(\bs{\beta}, \bs{z}^{(n)})=\sum_{j=1}^n w(r^2_j/c^*)r^2_j(\bs{\beta})&\geq w(r^2_{i_0}/c^*)r^2_{i_0}(\bs{\beta}) 
=r^2_{i_0}(\bs{\beta}) \nonumber\\[0ex]
&> \Big((1+\eta)\sqrt{M}-|y_{i_0}|\Big)^2  
\geq \Big((1+\eta)\sqrt{M}-\sqrt{M}\Big)^2 \nonumber\\[1ex]
&=\eta^2M 
>M \geq  O( \bs{0}_{p \times 1}, \bs{z}^{(n)}). \nonumber
\end{align}

\noin
That is, we have certified (\ref{proof.inequality}). \hfill \pend

\vs

\noin
\tb{Proof of Lemma 3.1}
Write $w(r^2/c^*)r^2=c^*w(r^2/c^*)r^2/c^*:=c^*w(x^2)x^2$, where $x=|r|/\sqrt{c^*}$ and $x^2=r^2/c^*>c$. It suffices to show that $w(x^2)x^2$ is 
strictly decreasing in $x$ (this intuitively is clear from Figure \ref{fig:w(r)r-sq}). Or equivalently, to show that the derivative of $w(x^2)x^2$ is negative. A straightforward calculus derivation yields
$$
\left(w(x^2)x^2\right)'=2x/(1-e^{-k})\left(e^{-k(1-c/x^2)^2} \big(1-2kc/x^2(1-c/x^2)\big)-e^{-k}\right). 
$$
Now it suffices to show that
\[
\left(e^{-k(1-c/x^2)^2} \big(1-2kc/x^2(1-c/x^2)\big)-e^{-k}\right)<0.
\]
Or equivalently to show that
\[
e^{k\left((1-c/x^2)^2-1\right)}>1-2kc/x^2(1-c/x^2). 
\]
For convenience, write  $t:=c/x^2$. Then $t\to 0$ as $x^2\to \infty$. Now we want to show that
\be
e^{-tk(2-t)}
> 1-2kt(1-t). \label{main.ineq}
\ee

A straightforward Taylor expansion of $e^x=1+x+x^2/2!+x^3/3!+ \cdots$  to the left hand side (LHS) of (\ref{main.ineq})  yields
\begin{align}
e^{-tk(2-t)}&=1+(-2kt+kt^2)+(-2kt+kt^2)^2/2+(-2kt+kt^2)^3/3!+(-2kt+kt^2)^4/4!+\cdots \nonumber\\[1ex]
& >1+(-2kt+kt^2)+(-2kt+kt^2)^2/2+(-2kt+kt^2)^3/3!\nonumber\\[1ex]
&=1-2kt(1-t)-kt^2+{(-kt(2-t))^2}/{6}+{2(-kt(2-t))^2}/{6}+{(-kt(2-t))^3}/{6} \nonumber\\[1ex]
&=1-2kt(1-t)+kt^2\Big({k(2-t)^2}/{6}-1\Big) +k^2t^2(2-t)^2\big(2-kt(2-t)\big)/6\nonumber\\[1ex]
&> 1-2kt(1-t)   \label{main.ineq2}
\end{align}
where the first inequality follows from the fact that
$$\frac{(-kt(2-t))^{2n}}{(2n)!}+\frac{(-kt(2-t))^{2n+1}}{(2n+1)!} 
=\frac{(-kt(2-t))^{2n}(2n+1-kt(2-t))}{(2n+1)!}>0,
$$
for $n \geq 2$ and small enough $t$.
And the last inequality in (\ref{main.ineq2}) follows the facts (i) $k(2-t)^2/6-1>0$ (if $t<2-\sqrt{6/k}$) and (ii) $2-kt(2-t)>0$ (if $t<1-\sqrt{1-2/k}$).
\vs
Combining (\ref{main.ineq2}) with (\ref{main.ineq}), we complete the proof.  \hfill \pend \vs

\vskip 3mm
\noindent
\tb{Proof of Theorem 3.3:}
It suffices to treat the case $p>1$ and furthermore by Theorem 4 on page 125 of \cite{RL87} 
it is sufficient to show that {$m=\lfloor{(n-p)}/{2}\rfloor$ contaminating points are not enough to break drown $\widehat{\bs{\beta}}_{wls}$.}
Assume it is otherwise. This  implies that either
\bi
\item[]\tb{(I)} $|\widehat{\bs{\beta}}^n_{wls}((\mb{z}_m^{(n)})_j)_1|\to \infty$ and $\|\widehat{\bs{\beta}}^n_{wls}((\mb{z}_m^{(n)})_j)_2\|$ is finite, or\vs
\item[]\tb{(II)}
$\|\widehat{\bs{\beta}}^n_{wls}(\mb{z}_m^{(n)})_j)_2\|=\big|\tan(\theta_{\widehat{\bs{\beta}}^n_{wls}(\mb{z}_m^{(n)})_j)})\big|\to \infty,$
\ei
along a sequence of $(\mb{z}_m^{(n)})_j$ as $j\to \infty$, where the subscripts $1$ and $2$ correspond to the intercept and non-intercept terms, respectively, as in the case $\bs{\beta}=(\beta_1, \bs{\beta}^{\top}_2)^{\top}$ in $\R^p$. We seek a contradiction for both cases. For description simplicity, write $\bs{\beta}_j:=\widehat{\bs{\beta}}^n_{wls}((\mb{z}_m^{(n)})_j)$
\vs
Case \tb{(I)}.  For simplicity, write $\bs{\beta}_j=(\beta_{1}, \bs{\beta}_{2}^{\top})^{\top}$
and $\bs{\beta}_{jj}=(2^m\beta_{1}, \bs{\beta}_{2}^{\top})^{\top}$. Then it is readily seen that
$r^2_i(\bs{\beta}_j)<r^2_i(\bs{\beta}_{jj})$ for large positive $m$. In light of Lemma 3.1, one has that
$O(\bs{\beta}_j)>O(\bs{\beta}_{jj})$, a contradiction is obtained.


\vs
Case \tb{(II)}. This case implies there is a sequence of hyperplanes induced from $ \widehat{\bs{\beta}}^n_{wls}((\mb{z}_m^{(n)})_j)$ that tend to the eventual vertical position as $j\to \infty$.
Denote by $H_j$ for those hyperplanes.  Let  $H_j$ intercept with the horizontal hyperplane $H_h$ at $\ell_j$, the hyperlines ( or the common part of $H_j$ and $H_h$). \vs
For simplicity, write the minimizer $\bs{\beta}_j=(\beta_1, \bs{\beta}^{\top}_2)^{\top}:=\widehat{\bs{\beta}}^n_{wls}((\mb{z}_m^{(n)})_j)$. Introduce a new hyperplane determined by $\bs{\beta}_{jj}=(\alpha\beta_1, \kappa\bs{\beta}^{\top}_2)^{\top}$ ($\kappa>1$ is a positive integer).
This $\bs{\beta}_{jj}$ amounts to tilting  $H_j$ (corresponding to $\bs{\beta}_j$) along $\ell_j$ to a more vertical position  $H_{jj}$ (corresponding to $\bs{\beta}_{jj}$). Note that it is possible that there is no data points touched during the titling process except those being originally on the $H_j$ since both hyperplanes are almost vertical. It is readily seen that $r^2_i(\bs{\beta}_{jj}) >r^2_i(\bs{\beta}_{j}) \to \infty$ except those points $(\bs{x}^{\top}_i, y_i))^{\top}$ that 
originally lie on the $\ell_{j}$ with zero residual. 
By Lemma 3.1, $O(\bs{\beta}_{j})>O(\bs{\beta}_{jj})$, a contradiction is reached. \hfill \pend
\vs
\end{document}